# Formation of Mn-rich interfacial phases in Co$_2$Fe$_x$Mn$_{1-x}$Si thin films

Ka Ming Law[1], Arashdeep S. Thind[2], Mihir Pendharkar[3], Sahil J. Patel[3], Joshua J. Phillips[1], Chris J. Palmstrom[3,4], Jaume Gazquez[5], Albina Borisevich[6], Rohan Mishra[7,2], and Adam J. Hauser[1]

[1]*Department of Physics and Astronomy, University of Alabama, Tuscaloosa, Alabama 35487, USA*

[2] *Institute of Materials Science & Engineering, Washington University in St. Louis, One Brookings Drive, St. Louis, Missouri 63130, USA*

[3]*Department of Electrical and Computer Engineering, University of California, Santa Barbara, Santa Barbara, California, 93106*

[4]*Materials Department, University of California, Santa Barbara, Santa Barbara, Santa Barbara, California, 93106*

[5] *Center for Nanophase Materials Sciences, Oak Ridge National Laboratory, Oak Ridge, Tennessee 37831, USA*

[6] *Institut de Ciència de Materials de Barcelona (ICMAB-CSIC), Campus UAB, Bellaterra 08193, Barcelona, Spain*

[7] *Department of Mechanical Engineering & Materials Science, Washington University in St. Louis, One Brookings Drive, St. Louis, Missouri 63130, USA*

**Abstract**

We report the formation of Mn-rich regions at the interface of $Co_2Fe_xMn_{1-x}Si$ thin films grown on GaAs substrates by molecular beam epitaxy (MBE). Scanning transmission electron microscopy (STEM) with electron energy loss (EEL) spectrum imaging reveals that each interfacial region: (1) is 1-2 nm wide, (2) occurs irrespective of the Fe/Mn composition ratio and in both Co-rich and Co-poor films, and (3) displaces both Co and Fe indiscriminately. We also observe a Mn-depleted region in each film directly above each Mn-rich interfacial layer, roughly 3 nm in width in the $x = 0$ and $x = 0.3$ films, and 1 nm in the $x = 0.7$ (less Mn) film. We posit that growth energetics favor Mn diffusion to the interface even when there is no significant Ga interdiffusion into the epitaxial film. Element-specific X-ray magnetic circular dichroism (XMCD) measurements show larger Co, Fe, and Mn orbital to spin magnetic moment ratios compared to bulk values across the $Co_2Fe_xMn_{1-x}Si$ compositional range. The values lie between reported values for pure bulk and nanostructured Co, Fe, and Mn materials, corroborating the non-uniform, layered nature of the material on the nanoscale. Finally, SQUID magnetometry demonstrates that the films deviate from the Slater-Pauling rule for uniform films of both the expected and the measured composition. The results inform a need for care and increased scrutiny when forming Mn-based magnetic thin films on III-V semiconductors like GaAs, particularly when films are on the order of 5 nm or when interface composition is critical to spin transport or other device applications.

**Introduction**

The Heusler material class has garnered significant attention in the last couple decades. Since the Heusler alloy $Cu_2MnAl$ has been shown to feature emergent magnetism and half-metallicity[1], a promising subclass, Co-based tertiary Heusler compounds of the form $Co_2YZ$, has been receiving intensive interest[2-6]. Among the choices of compounds, $Co_2FeSi$ and $Co_2MnSi$ are predicted to be half-metallic, and bulk studies have demonstrated high Curie temperatures of over 900 K[5, 7]. As such, they become ideal candidates for nonvolatile magnetic memory applications. Substantial work has been done with $Co_2FeSi$ and $Co_2MnSi$ as electrodes for magnetic tunneling junctions[8-14], and in recent years for spin injectors[15-18].

$Co_2FeSi$ and $Co_2MnSi$ are isomorphic and have nearly identical bulk lattice constants of 5.654[7] and 5.658 Å, respectively[19]. If one considers the two materials as two endpoints of a series of continuously varying Fe:Mn ratio, it stands to reason that the high-degree of structural similarity between the endpoint compounds should also apply to any point along the series, i.e., $Co_2Fe_xMn_{(1-x)}Si$. The implication is the prospect of tuning the magnetic properties through Fe:Mn stoichiometry in epitaxial $Co_2Fe_xMn_{(1-x)}Si$ thin films. However, such tuning requires understanding of how the bulk film and interfaces are affected by their environment, including the choice of the substrate.

In this Article, we report on the properties of ~5 nm thick $Co_2Fe_xMn_{1-x}Si$ thin films grown on GaAs substrates by molecular beam epitaxy (MBE). A 1-2 nm thick region with increased Mn content (at the equal cost of Co and Fe) appears to be energetically preferred at the interface. As the interfacial layers (and the 1-3 nm Mn-depleted region above) are a significant fraction of the ultrathin film, each film can now be viewed as several nanoscale layers of different compositions. Not surprisingly, the magnetic properties of each film are significantly impacted: The Slater-Pauling rule for the expected compositions no longer match the experimentally determined saturation magnetization at T = 10K, and x-ray magnetic circular dichroism measurements show that the Co, Fe, and Mn films behave closer to nanomaterials than bulk materials, with orbital moments enhanced relative to spin counterparts.

**Methodology**

Co$_2$Fe$_x$Mn$_{(1-x)}$Si thin films on GaAs (001) substrates were grown by MBE. The three samples in this study were all grown at a substrate temperature of 800 °C and were made with Fe content of x = 0, 0.3, and 0.7. The composition and crystal quality of all thin films were evaluated using aberration-corrected scanning transmission electron microscopy (STEM). The STEM experiments were performed using two microscopes: a JEOL ARM200c and a Nion UltraSTEM 200. Both the STEMs were equipped with a fifth-order aberration corrector and a cold-field emission electron gun and electron energy loss spectroscopy (EELS) capabilities. The electron transparent samples were prepared by parallel polishing cross-section samples, followed by low angle ion milling at 3 kV followed by final polishing at 1 kV. The atomic resolution high-angle annular dark-field (HAADF) images and EELS composition profiles for Co L, Mn L, and Fe L (where appropriate) edges are shown in **Figure 1**. Si maps were not obtained due to the proximity of the Ga M and Si L edges, and poor signal-to-noise ratio of Si K edge.

We performed XMCD measurements on the three samples at the (now decommissioned) 4-ID-C beamline in 2017. Total fluorescence yield (TFY) x-ray absorption spectra were collected at the L$_{2,3}$ edges of Co (770-810 eV), Mn (630-670 eV), and Fe (700-740 eV), under an external field of ±1 T and sample temperature of 20 K. The external field and incident beam are parallel and at an angle of 20° above the sample surface. External magnetic field switching occurs after all samples have been measured in order to minimize the number of times of switching during a beamline session. During each measurement, the incident beam polarization is toggled between right-circular (RCP) and left-circular (LCP) at each energy sampling step, producing two absorption spectra at each field value, totaling four spectra for each elemental L-edge. Each spectrum is normalized by its right (high-energy) step edges.

The difference spectrum $\mu_{RCP} - \mu_{LCP}$ is known as the circular dichroism spectrum, and scales with the angle between magnetization and polarization vector (using right/left-hand rule) when the sample is under magnetic saturation; with the magnetization and polarization along the same axis, as was the case with our measurement geometry, reversing simultaneously the field direction and polarization produces the same absorption spectrum[20, 21]. This phenomenon is widely exploited by the community to further improve the signal-to-noise level. We verified that above symmetry was indeed present among our spectra, then performed the following calculations:

$$\mu_+ = (\mu_{RCP,+1T} + \mu_{LCP,-1T})/2$$

$$\mu_- = (\mu_{LCP,+1T} + \mu_{RCP,-1T})/2,$$

where $\mu_+$ and $\mu_-$ denote the absorption spectra with the magnetization parallel and antiparallel to the polarization, respectively. We then refer $\mu_+ - \mu_-$ as the magnetic dichroism (XMCD) and $\mu_+ + \mu_-$ as the total x-ray absorption (XAS).

For an atom with a partially-filled outer shell, the integrated intensities of the total XAS and the XMCD at the $L_2$ and $L_3$ edge can be used to calculate the average spin-[22] and orbital-[23] angular momentum. These are called the magneto-optical sum rules:

$$\frac{\int_{L_3} dE \, (\mu_+ - \mu_-) - 2\int_{L_2} dE \, (\mu_+ - \mu_-)}{\int dE \, (\mu_+ + \mu_-)} = \frac{1}{n_h}\langle S_z \rangle$$

$$\frac{\int_{L_3} dE \, (\mu_+ - \mu_-) + \int_{L_2} dE \, (\mu_+ - \mu_-)}{\int dE \, (\mu_+ + \mu_-)} = \frac{3}{4n_h}\langle L_z \rangle,$$

as lifted from their respective references, where $c = 1$, $l = 2$, and $(4l + 2 - n) = n_h$ is the number of holes in the d-shell (as stated in [23]). The isotropic (unpolarized) total x-ray absorption integrated intensity is assumed as $\mu_0 = (\mu_+ + \mu_-)/2$. [24, 25]

It is well-known within the community that the choice of $n_h$ is highly non-trivial: the occupancy of the outer valence shell depends heavily on the lattice structure and the composition of the sample[26] and the presence of saturation effects (regarding electronic excitations)[27]. To avoid the technicality and even potential subjectivity, the ratio of spin to orbital moment is usually reported instead, in the literature:

$$\frac{\frac{1}{n_h}\langle S_z\rangle}{\frac{1}{n_h}\langle L_z\rangle} = \frac{\langle S_z\rangle}{\langle L_z\rangle} = \frac{1}{2}\frac{m_s}{m_l}.$$

In order to calculate the total XAS integrated intensities, a multi-step-edge baseline needs to be removed from each of $\mu_+$ and $\mu_-$. We subtract an arctangent function at each L-edge, with the ratio of the $L_2$ and $L_3$ step heights equal to the ratio of the tallest absorption peak heights [28-30].

Magnetic properties were measured using a Quantum Design Magnetic Property Measurement System (MPMS-3) utilizing a superconducting quantum interference device (SQUID). Magnetic hysteresis measured in the in-plane geometry at 10 K, to a maximum external applied field magnitude of 9 T.

**Results and Discussion**

*Scanning transmission electron microscopy (STEM).* HAADF STEM images are shown in **Figures 1a, 1b, and 1c**, and EELS composition profiles as a function of film depth for Co, Mn and Fe are shown in **Figures 1d, 1e, and 1f** for intended compositions x = 0, 0.3, and 0.7, respectively. We noted the uncertainty of the abruptness of the interface between the top of the film and the capping layer, due to both sample tilt and potential surface roughness. Thus, we marked in **Figures 1d** (x = 0) and **1e** (x = 0.3) a reasonable estimate wherein capping layer and film coexist through the cross-section, and we did not attempt to analyze this region. In **Figure 1f**, the contact layer appears epitaxial resulting in a sharp interface between the film and the capping layer.

We estimated the compositions of the Co-Fe-Mn sublattice of each film by integrating the EEL spectra over the range of the green rectangle shown in each HAADF STEM image for Co L, Fe L and Mn L edges. We extracted the compositions of $Co_{2.06}Mn_{0.94}$, $Co_{1.82}Fe_{0.46}Mn_{0.72}$ and $Co_{1.35}Fe_{1.4}Mn_{0.24}$ for x = 0, 0.3, and 0.7, respectively. Due to the proximity of the Ga M and Si L edges, and poor signal-to-noise ratio of the Si K edge, we cannot unambiguously quantify the Si concentration. We note that while the x = 0 sample is slightly Co-rich, the x = 0.3 and x = 0.7 samples are significantly Co-poor, but still maintain what appear to be B2 and $L2_1$-like phases as shown in **Figures 1b and 1c**. Discussion of the phase separation can be found later in the section.

The HAADF image in **Figure 1a** of $Co_2MnSi$ (x=0) displays uniform B2 ordering: Mn and Si atoms are seen to be uniformly mixed within a sublattice of Co atoms. **Figure 1d** shows EELS elemental line profile across the thickness of the x = 0 sample. While the overall composition and surface (near film depth value of 0) stoichiometry are close to the expected values, the film exhibits gradients in composition near the substrate/film interface. From the film surface towards the substrate, increasing Co concentration substitutes for Mn sites until a depth of 2-3 nm, where the composition is $Co_{2.28}Mn_{0.72}Si$. Within the remaining ~2 nm of the film, Mn:Co shifts to become increasingly Mn-rich, with a interfacial composition of $Co_{1.5}Mn_{1.5}Si$.

We surmise that the replacement of Co with Mn in the lower half of the film is caused by Mn preferentially migrating towards the GaAs substrate during the deposition process to form MnGaAs phases that are more energetically-favorable—the diffusion of Mn into GaAs to form multiphase Mn-Ga-As is well-documented in studies of Mn thin films on GaAs substrates[31, 32]. Since the overall (integrated) composition of the film is close to expected ($Co_{2.06}Mn_{0.94}$), the accumulated Mn at the substrate-film interface appears to originate from the "Mn-depleted" region located 1-3.5 nm from the film surface (**Figure 1**); the drop in Mn concentration in the region was then subsequently filled with Co displaced from the interface.

Singh et al.[33] found that 300 nm $Co_2MnSi$ films grown by on-axis DC magnetron sputtering at 425°C yielded a distinct interfacial Mn-As epitaxial layer with no Co, followed by a Co-Ga rich region further into the film, a sign of deep Ga diffusion in their films. Since Si L and Ga M edges overlap heavily in EELS spectra it would be challenging to pick out any qualitative hints of Ga content in our MBE-grown films, but **Figures 1d-f** show that Mn remains less than 50% of the interfacial composition. This contrasts with Singh et al. and suggests that if interdiffusion is occurring, it is likely not significant enough to cause a Mn-As phase in our films. However, our results match closely a wide swath of results that suggest Mn energetically prefers to aggregate in films near the interface of GaAs substrate[34-36]. First-principles calculations of $Co_2MnSi$/GaAs have shown that, on a GaAs(100)-textured surface, the CoMn-terminated surface is more energetically favorable than the CoSi-terminated surface[37], and that terminations containing Mn are generally more stable when Co and Mn atoms have high chemical potential[38].

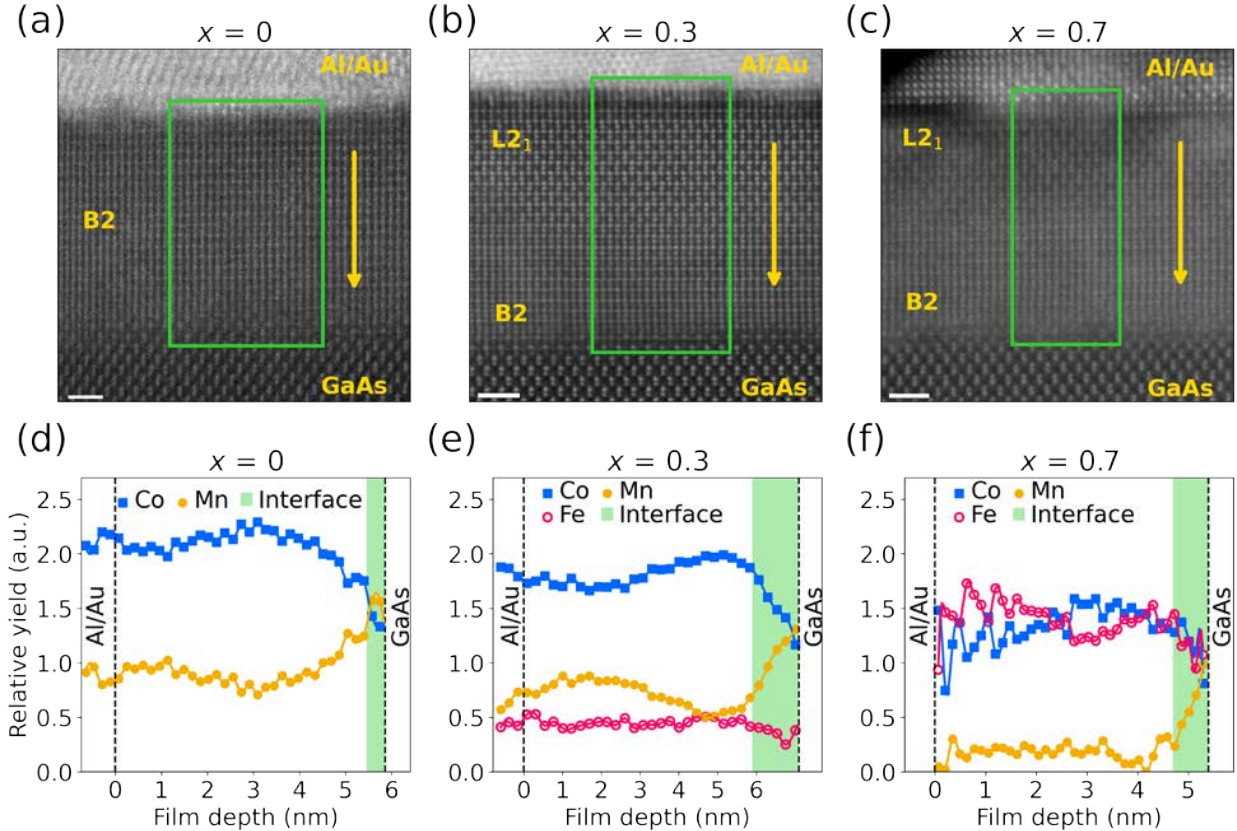

Figure 1: STEM HAADF images showing the cross-section of the $Co_2Fe_xMn_{(1-x)}Si$ thin films on GaAs (001) substrate for (a) x=0, (b) x=0.3 and (c) x=0.7 samples. Scale bars correspond to 1nm. The EELS composition profiles showing the variation of Co, Fe and Mn across the film depth for (d) x=0, (e) x=0.3 and (f) x=0.7. The EELS data was acquired for the region marked as the green box in the corresponding HAADF images for x=0, x=0.3 and x=0.7. The arrows indicate the direction of the EELS profile across the film depth (from contact towards the substrate).

The EELS composition profiles as a function of film depth for films grown with Fe fractions x = 0.3 (**Figure 1e**), and x = 0.7 (**Figure 1f**) show the same qualitative pattern as the x = 0 sample. In both cases, we see a large increase in Mn fractional share of the composition in the 1-2 nm near the substrate/film interface, and a subsequent Mn-depleted region a few nanometers wide as one looks immediately further from the interface. The Mn-rich interfacial state appears to replace both Co and Fe at similar proportional fractions, and occurs regardless of Mn/Fe content ratios or whether the film is slightly Co-rich overall (x = 0) or Co-deficient (as in the case of the x = 0.3 and 0.7 samples).

HAADF images also reveal that there is an abrupt structural phase transition in the two samples containing Fe (**Figure 1b**, **1c**). In both cases, the top layer has L2$_1$ ordering, while the bottom layer has B2 ordering. The vertical location of the transition coincides closely with the top of the Mn-depleted region in both films. In other words, we observe B2 ordering in the regions where Mn segregation is causing significant fluctuations in the composition of the material, and L2$_1$ ordering when the shifts are less severe. In the x = 0 film (**Figure 1a**), we only see B2 phase and this may be due to significant changes in the composition (both Co and Mn) nearly all the way to the top of the film. However, whether compositional changes above the Mn-depleted regions in the x = 0 sample are indeed smaller in magnitude than in the x = 0.3 and 0.7 compositions is debatable. We also cannot easily attribute the L2$_1$-like ordering to a steady Fe content in those regions, as the Fe content in the x = 0.7 film (**Figure 1f**) is either highly variable or highly uncertain from film depths of 0-2 nm. It is noteworthy that the films that are Co-poor show L2$_1$ ordering, but the film that has a Co/(Fe+Mn) ratio close to 2 show no observed L2$_1$ ordering in the region of the sample characterized. The driver of ordering herein thus remains an open question, though double substitutions have been proposed wherein Fe preferentially sits on the Co sites when displaced by Mn or when there are too few Co atoms to maintain the epitaxial structure alone[39, 40].

*X-ray absorption spectroscopy.* The spectra $\mu_+$ and $\mu_-$, as well as total XAS and XMCD spectra, collected at the L$_{2,3}$-edge of each element (Co, Mn, and Fe) are presented in **Figure 2**, **Figure 3**, and **Figure 4**, respectively. The x = 0 sample does not have data at the Fe L$_{2,3}$-edge (**Figure 4**). The spin and orbital magnetic moment per valence-shell hole values are listed in **Table 1**, and their ratios are listed in **Table 2**. We observed that for the x = 0 sample, our $m_l/m_s$ values for Co and Mn are drastically larger than those of previous Co$_2$MnSi work. One work[41] reported (for Co and Mn, respectively) 0.045 and 0.015, and another[42] reported 0.050 and 0.022. Even though the x-ray absorption spectra for these works are sampled at room temperature (and ours sampled at 20K), the (1) high Curie temperature of Co$_2$MnSi and (2) the fact that sub-RT magnetic transitions have not been seen[43], points to a different origin for these mismatches. Early XMCD work on Co thin films reported on a $m_l/m_s$ value of 0.095[28], and 0.24 for Co nanoparticles[44]. We posit that the departure from Heusler values in x = 0 of **Table 2** is a result of the compositional segregation, creating nanoscale structures with increased concentration of Co-Co bonding, and thus an increase in the orbital moment.

Table 1: Orbital and spin magnetic moment per hole. Conservative estimate of uncertainty, based on the center positions of the baseline step-edges relative to the white line, is ±15% for orbital moment per hole and ±5% for spin moment per hole.

| X = | | Magnetic moment per hole ($\mu_B/n_h$) | | |
|---|---|---|---|---|
| | | Co site | Mn site | Fe site |
| 0 | Orb | 0.05(1) | 0.13(2) | |
| | Spin | 0.39(2) | 0.46(2) | |
| 0.3 | Orb | 0.06(1) | 0.14(2) | 0.04(1) |
| | Spin | 0.48(2) | 0.61(3) | 0.71(4) |
| 0.7 | Orb | 0.05(1) | 0.15(2) | 0.10(2) |
| | Spin | 0.51(3) | 0.58(3) | 0.66(3) |

Table 2: Ratio between orbital and spin magnetic moment. Absolute uncertainties propagated from values in Table 1.

| X = | $m_l/m_s$ | | |
|---|---|---|---|
| | Co site | Mn site | Fe site |
| 0 | 0.13(2) | 0.28(4) | |
| 0.3 | 0.13(2) | 0.24(3) | 0.06(2) |
| 0.7 | 0.10(2) | 0.26(4) | 0.15(3) |

The calculated spin moment per hole of Co, calculated from the Co $L_{2,3}$ edge, increases by a statistically significant amount with x (**Table 1**), even when considering the 5% uncertainty assigned to each value. A significant orbital moment contribution per hole is also observed, remaining constant within our estimated measurement uncertainty of 15%. While these results might suggest that the orbital-to-spin moment ratios in **Table 2** would be decreasing with x, the propagated uncertainties for the ratios are such that the ratios for all compositions sit at a constant value within the experimental uncertainty. This likewise occurs across the series for Mn sites. Thus, for both Co and Mn sites, it is reasonable to conclude that $m_l/m_s$ values remain constant across the series.

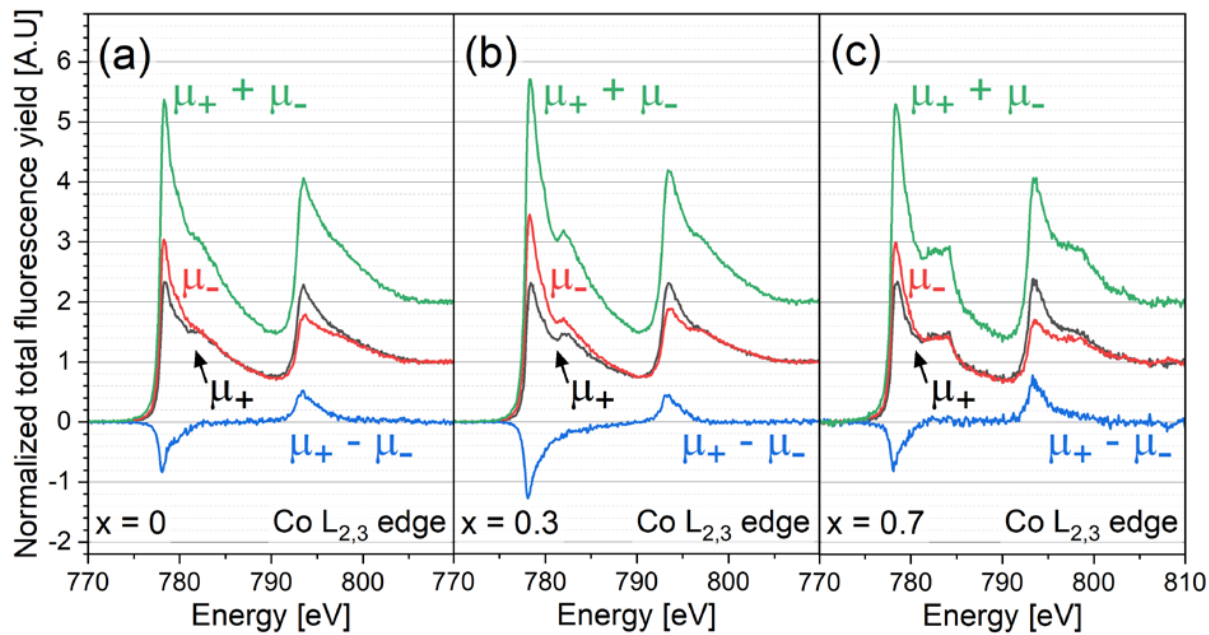

Figure 2: XAS and XMCD at Co $L_{2,3}$-edge for the x=0 sample (a), x=0.3 sample (b), and x=0.7 sample (c).

The XAS and XMCD at the Co $L_{2,3}$ edge (**Figure 2**) are qualitatively identical across all samples, other than a secondary peak feature that gradually emerges to the right side of the main peak at 784.2 eV (798.6 eV) at the $L_3$ ($L_2$) edge with increasing x. This secondary peak feature is most prominent in the x=0.7 sample, and has a non-negligible dichroic response in both the $L_2$ and $L_3$ edge. They do not appear to be a part of the multiplet fine structure of Co x-ray absorption, and there are only rare mentions of it among literatures. An in-depth study of this feature, which were found in the x-ray absorption of Co nanocrystals suspended in liquids[45], concluded that it is due to metal-to-ligand charge transfer; however, our thin film samples were deposited by MBE and were not suspended in any liquids. An XAS study on Co polyoxometalates found the same feature in their sample, but they were unable to reproduce it through multiplet calculations[46]. A peak at 783.5-784.0 eV can be found in the work[47]; in this work, authors calculated the $L_3$ edge of $Co^{2+}$ and $Co^{3+}$ in the presence of an octahedral crystal field. However, our main peak of 778 eV is not present in their calculated Co XAS spectra. The authors of Ref. [48] observed this feature in the Co $L_3$ XMCD of a different Heusler system, and made a postulate based on band structure calculations they reported in the same work.

In the absorption spectra at the Mn $L_{2,3}$ edge, multiplet fine structures are likewise not observed. As is with Co $L_{2,3}$ edge, the total XAS do not appear to follow a monotonic trend from x = 0 to 0.3 to 0.7. On the other hand, the XMCD signal at the Mn $L_3$ edge clearly increases in intensity across the samples. We also note that, within the XMCD spectrum of each sample (**Figure 3a**, **Figure 3b**), a strong positive shoulder is observed to the right side of the $L_3$ main peak; noting that the Mn $L_3$ main peak has negative intensity, the shoulder thus leads to a significant suppression of the Mn spin and orbital moment per hole values (**Table 1**).

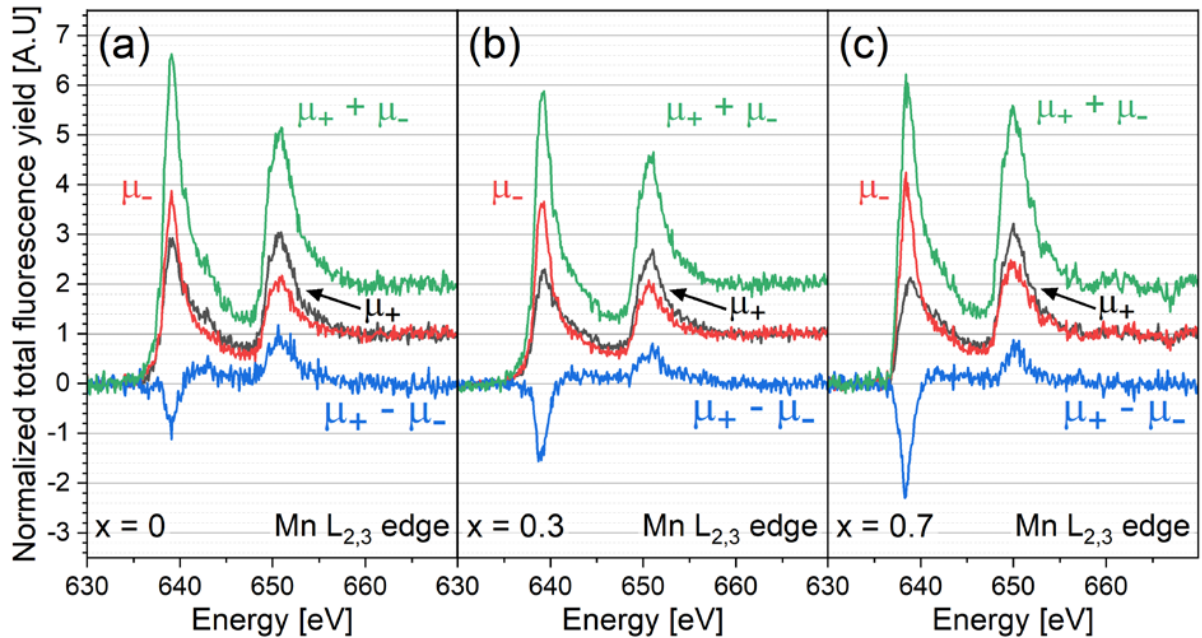

Figure 3: XAS and XMCD at Mn $L_{2,3}$-edge for the x=0 sample (a), x=0.3 sample (b), and x=0.7 sample (c).

Lastly, **Figure 4a** and **4b** shows the total XAS and XMCD at the Fe $L_{2,3}$-edge in the x = 0.3 and 0.7 samples. Notably, the $L_2$ main peak has a larger intensity than the $L_3$ main peak in $\mu_+$ of either case. The strong XMCD response yields the largest spin-angular momentum per hole for Fe, relative to those calculated for Co and Mn (**Table 1**). As with the Mn $L_{2,3}$-edge (**Figure 3a**, **3b**), no multiplet fine structures are observed, and positive shoulders near the $L_3$ main peaks are small.

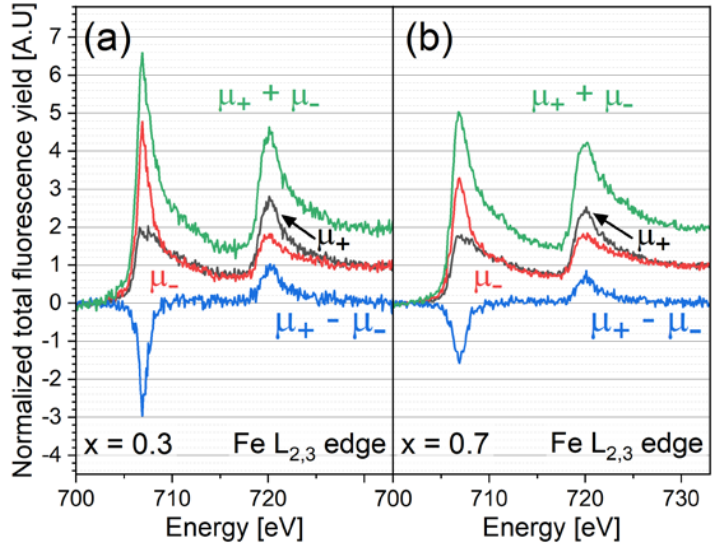

Figure 4: XAS and XMCD at Fe $L_{2,3}$-edge for the x=0.3 sample (a) and x=0.7 sample (b).

*Magnetic Characterization.* The expected magnitude of the saturation magnetization *M* of a material can be estimated as a function of the total number of valence electrons *N* using the $M = N - 24$ generalized Slater-Pauling rule[6], assuming a uniform composition. Although we have proven above that the films are not uniform, we corroborate this finding using SQUID magnetometry. Magnetization vs applied magnetic field (M-H) curves taken at T = 10 K are shown for x = 0 (black), 0.3 (red) and 0.7 (blue) in **Figure 5**, with an inset showing the low field hysteresis of each system. All samples were measured starting from H = +9 T and swept to H = -9 T and back in each curve, with no significant increase in moment above a magnitude of 1 T. The magnetization of each sample is based on the thickness directly measured in **Figure 1**, and the atomic spacing in each sample directly measured to ensure that the estimated unit-cell volume of 181 Å$^3$ is appropriate.

If one uses the overall compositions calculated from EELS data by integrating all signal from the green box in **Figures 1a-c**, we arrive at the rather constant values of 5.12, 5.10, and 5.03 $\mu_B$ per formula unit (f.u.) for x = 0, 0.3, and 0.7, respectively, in contrast to the expected values of 5.0, 5.3, and 5.7 $\mu_B$/f.u. (also respectively). However, the magnetometry data for our samples (**Figure 5**) suggests a larger variation among saturation moments for the samples, with values of 4.7, 5.3, and 6.4 $\mu_B$ per formula unit (again, respectively). Even should we assign exceptionally large uncertainty values (e.g. ±0.2 $\mu_B$/f.u.) to our measurements, the films can scarcely be said to fit either the naïve x values based on the assumed compositions, and are nowhere near the ~5 $\mu_B$/f.u. values that the EELS-derived compositions would give. We cannot expect any significant moments from the Si sites, and even if there were significant moments, the total magnetic moment would be moved *further* from the Slater-Pauling values, not closer.

It is worth noting that the magnitude of the ferromagnetic hysteresis alone (**Figure 5 *Inset***) are quite comparable to each other, and the difference in saturation magnetization appears to stem from a superparamagnetic contribution that contributes moment increasingly as the Fe content increases. This is also consistent with EELS and XMCD data that point to nanostructured elements, as it is widely documented that nanoparticles of and discontinuous film regions in Co, Fe, Mn and many alloys thereof (including CFMS and its parent compounds $Co_2FeSi$ and $Co_2MnSi$) can show significant levels of superparamagnetism[49-57].

Overall, the results point to the need for close scrutiny of microstructure when encountering non-Slater-Pauling results of any kinds in this and similar alloys, as it may be a warning bell of film non-uniformity just as often as a sign of exotic physics. We further suggest scrutiny for those samples that show evidence of superparamagnetic components within the material.

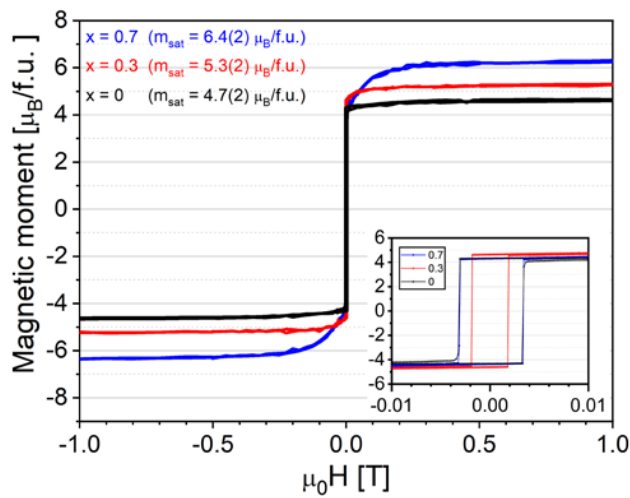

Figure 5: Magnetization vs. applied Magnetic Field (M-H) curves determined via SQUID magnetometry. *Inset:* M-H curves zoomed in to show low-field hysteresis for x = 0 (black hollow diamonds), 0.3 (red hollow diamonds) and 0.7 (blue filled circles) samples.

**Conclusions**

We have revealed that each of our films features a highly nontrivial multilayer structure with both compositional and structural phase segregation, and that the interfacial stoichiometry may sometimes hugely deviate from the average stoichiometry of the thin film. We note that Mn accumulation at the substrate-film interface is observed regardless of the expected and overall composition. In ultrathin Co-based Heusler films with thickness comparable to 5 nm, the migration of Mn will cause significant compositional deviations across the film, and the stoichiometry at the film surface may not be guaranteed. This poses a problem for applications that are sensitive to interfacial magnetic properties such as spin valves and magnetic tunnel junctions. The bulk magnetic properties of our ultrathin film are affected as well, as saturation moment and $m_l/m_s$ values similarly deviate from what is expected from published work on Heusler alloys.

Based on our findings, we suspect that the tendency for these elements to separate into energetically preferable Heusler structures may explain non-monotonic (average) compositional- and thickness-dependence in the magnetoresistance ratio (MR) of magnetic tunnel-junctions made out of Co-based Heusler thin films[14]—although, in the work cited, the Heusler layer is deposited on MgO instead of on GaAs as in this work, and more studies still need to be done regarding compositional and phase segregation of $Co_2Fe_xMn_{(1-x)}Si$ deposited on other substrates.

**Acknowledgements**

We acknowledge support from the National Science Foundation (NSF), USA through NSF-CAREER Award No. DMR-2047251. ICTS-CNME at UCM is acknowledged for offering access to STEM microscopy and expertise. ICMAB author acknowledges Spain's Agencia Estatal de Investigación Severo Ochoa Program for Centers of Excellence in R&D (CEX2019-000917-S). The work at Washington University was supported by the NSF through awards # DMR-1806147 and DMR-2145797. A portion of the STEM experiments were performed at the Center for Nanophase Materials Sciences (CNMS), which is a US Department of Energy, Office of Science User Facility at Oak Ridge National Laboratory.

**Declaration of Competing Interest**

The authors declare that they have no known competing financial interests or personal relationships that could have appeared to influence the work reported in this paper.

# References

1. de Groot, R.A., et al., *New Class of Materials: Half-Metallic Ferromagnets.* Physical Review Letters, 1983. **50**(25): p. 2024-2027.
2. Picozzi, S., A. Continenza, and A.J. Freeman, *Co2MnX(X=Si,Ge, Sn) Heusler compounds: An ab-initio study of their structural, electronic, and magnetic properties at zero and elevated pressure.* Physical Review B, 2002. **66**(9).
3. Sargolzaei, M., et al., *Spin and orbital magnetism in full Heusler alloys: A density functional theory study ofCo2YZ(Y=Mn,Fe;Z=Al,Si,Ga,Ge).* Physical Review B, 2006. **74**(22).
4. Kandpal, H.C., et al., *Correlation in the transition-metal-based Heusler compounds Co2MnSi and Co2FeSi.* Physical Review B, 2006. **73**(9).
5. Wurmehl, S., et al., *Geometric, electronic, and magnetic structure ofCo2FeSi: Curie temperature and magnetic moment measurements and calculations.* Physical Review B, 2005. **72**(18).
6. Galanakis, I., P.H. Dederichs, and N. Papanikolaou, *Slater-Pauling behavior and origin of the half-metallicity of the full-Heusler alloys.* Physical Review B, 2002. **66**(17).
7. Webster, P.J., *Magnetic and chemical order in Heusler alloys containing cobalt and manganese.* Journal of Physics and Chemistry of Solids, 1971. **32**(6): p. 1221-1231.
8. Hülsen, B., M. Scheffler, and P. Kratzer, *Structural Stability and Magnetic and Electronic Properties of Co2MnSi(001)/MgO Heterostructures: A Density-Functional Theory Study.* Physical Review Letters, 2009. **103**(4).
9. Gercsi, Z., et al., *Spin polarization of Co2FeSi full-Heusler alloy and tunneling magnetoresistance of its magnetic tunneling junctions.* Applied Physics Letters, 2006. **89**(8).
10. Tsunegi, S., et al., *Large tunnel magnetoresistance in magnetic tunnel junctions using a Co2MnSi Heusler alloy electrode and a MgO barrier.* Applied Physics Letters, 2008. **93**(11).
11. Schmalhorst, J., et al., *Chemical and Magnetic Interface Properties of Tunnel Junctions With Co2MnSi/Co2FeSi Multilayer Electrode Showing Large Tunneling Magnetoresistance.* IEEE Transactions on Magnetics, 2007. **43**(6): p. 2806-2808.
12. Miura, Y., et al., *Coherent tunnelling conductance in magnetic tunnel junctions of half-metallic full Heusler alloys with MgO barriers.* Journal of Physics: Condensed Matter, 2007. **19**(36).
13. Liu, H.-x., et al., *Giant tunneling magnetoresistance in epitaxial Co2MnSi/MgO/Co2MnSi magnetic tunnel junctions by half-metallicity of Co2MnSi and coherent tunneling.* Applied Physics Letters, 2012. **101**(13).
14. Sakuraba, Y., et al., *Extensive study of giant magnetoresistance properties in half-metallic Co2(Fe,Mn)Si-based devices.* Applied Physics Letters, 2012. **101**(25).
15. Pfeiffer, A., et al., *Spin currents injected electrically and thermally from highly spin polarized Co2MnSi.* Applied Physics Letters, 2015. **107**(8).
16. Kimura, T., et al., *Erratum: Room-temperature generation of giant pure spin currents using epitaxial Co2FeSi spin injectors.* NPG Asia Materials, 2012. **4**(3): p. e13-e13.
17. Bruski, P., et al., *All-electrical spin injection and detection in the Co2FeSi/GaAs hybrid system in the local and non-local configuration.* Applied Physics Letters, 2013. **103**(5).
18. Hamaya, K., et al., *Estimation of the spin polarization for Heusler-compound thin films by means of nonlocal spin-valve measurements: Comparison of Co2FeSi and Fe3Si.* Physical Review B, 2012. **85**(10).
19. Niculescu, V., et al., *Relating structural, magnetic-moment, and hyperfine-field behavior to a local-environment model inFe3−xCoxSi.* Physical Review B, 1979. **19**(1): p. 452-464.
20. O'Brien, W.L. and B.P. Tonner, *Orbital and spin sum rules in x-ray magnetic circular dichroism.* Physical Review B, 1994. **50**(17): p. 12672-12681.

21. Rogalev, A., et al., *X-ray Magnetic Circular Dichroism: Historical Perspective and Recent Highlights*, in *Magnetism: A Synchrotron Radiation Approach*. 2006. p. 71-93.
22. Carra, P., et al., *X-ray circular dichroism and local magnetic fields.* Physical Review Letters, 1993. **70**(5): p. 694-697.
23. Thole, B.T., et al., *X-ray circular dichroism as a probe of orbital magnetization.* Physical Review Letters, 1992. **68**(12): p. 1943-1946.
24. Stöhr, J., *Exploring the microscopic origin of magnetic anisotropies with X-ray magnetic circular dichroism (XMCD) spectroscopy.* Journal of Magnetism and Magnetic Materials, 1999. **200**(1-3): p. 470-497.
25. Stöhr, J., *Quantum Theory of X-Ray Dichroism*, in *The Nature of X-Rays and Their Interactions with Matter*. 2023, Springer. p. 537-593.
26. Yonamoto, Y., et al., *Magnetism of an ultrathin Mn film on Co(100) and the effect of oxidation studied by x-ray magnetic circular dichroism.* Physical Review B, 2001. **63**(21).
27. Nakajima, R., J. Stöhr, and Y.U. Idzerda, *Electron-yield saturation effects in L-edge x-ray magnetic circular dichroism spectra of Fe, Co, and Ni.* Physical Review B, 1999. **59**(9): p. 6421-6429.
28. Chen, C.T., et al., *Experimental Confirmation of the X-Ray Magnetic Circular Dichroism Sum Rules for Iron and Cobalt.* Physical Review Letters, 1995. **75**(1): p. 152-155.
29. Cramer, S.P. and S.P. Cramer, *XANES and XMCD.* X-Ray Spectroscopy with Synchrotron Radiation: Fundamentals and Applications, 2020: p. 165-190.
30. Wilhelm, F., J. Sanchez, and A. Rogalev, *Magnetism of uranium compounds probed by XMCD spectroscopy.* Journal of Physics D: Applied Physics, 2018. **51**(33): p. 333001.
31. Osafune, Y., et al., *Depth profile photoemission study of thermally diffused Mn/GaAs (001) interfaces.* Journal of Applied Physics, 2008. **103**(10).
32. Hilton, J.L., et al., *Interfacial reactions of Mn/GaAs thin films.* Applied Physics Letters, 2004. **84**(16): p. 3145-3147.
33. Singh, L.J., et al., *Interface effects in highly oriented films of the Heusler alloy Co2MnSi on GaAs(001).* Journal of Applied Physics, 2006. **99**(1).
34. Zhang, M., G. Dong, and X. Zhu. *Formation of the Meta-Stable γ-Mn and GaAs (100) Interface: Diffusion and Chemical Reaction*. in *Chinese Science Abstracts Series A*. 1995.
35. Moiseev, K.D., et al., *On the delta-type doping of GaAs-based heterostructures with manganese compounds.* Semiconductors, 2017. **51**: p. 1141-1147.
36. Tansley, T. and P. Newman, *Measurements of heterojunctions alloyed on to GaAs.* Solid-State Electronics, 1967. **10**(5): p. 497-501.
37. Chen, L.-Y., et al., *Ab initio calculation of Co2MnSi/semiconductor (SC=GaAs, Ge) heterostructures.* Thin Solid Films, 2011. **519**(13): p. 4400-4408.
38. Hashemifar, S.J., P. Kratzer, and M. Scheffler, *Preserving the Half-Metallicity at the Heusler Alloy Co2MnSi(001) Surface: A Density Functional Theory Study.* Physical Review Letters, 2005. **94**(9).
39. Mahat, R., et al., *Effect of mixing the low-valence transition metal atoms Y= Co, Fe, Mn, Cr, V, Ti, or Sc on the properties of quaternary Heusler compounds Co 2− x Y x FeSi (0 ⩽ x ⩽ 1).* Physical Review Materials, 2022. **6**(6): p. 064413.
40. Shambhu, K., et al., *Co2Fe1. 25Ge0. 75: A single-phase full Heusler alloy with highest magnetic moment and Curie temperature.* Acta Materialia, 2022. **236**: p. 118112.
41. Saito, T., et al., *X-ray absorption spectroscopy and x-ray magnetic circular dichroism of epitaxially grown Heusler alloy Co2MnSi ultrathin films facing a MgO barrier.* Applied Physics Letters, 2007. **91**(26).
42. Stadler, S., et al., *Element-specific magnetic properties of Co2MnSi thin films.* Journal of Applied Physics, 2005. **97**(10).

43. Ritchie, L., et al., *Magnetic, structural, and transport properties of the Heusler alloys Co2MnSi and NiMnSb.* Physical Review B, 2003. **68**(10).
44. Wiedwald, U., et al., *Ratio of orbital-to-spin magnetic moment in Co core-shell nanoparticles.* Physical Review B, 2003. **68**(6).
45. Liu, H., et al., *Electronic Structure of Cobalt Nanocrystals Suspended in Liquid.* Nano Letters, 2007. **7**(7): p. 1919-1922.
46. Hibberd, A.M., et al., *Co Polyoxometalates and a Co3O4 Thin Film Investigated by L-Edge X-ray Absorption Spectroscopy.* The Journal of Physical Chemistry C, 2015. **119**(8): p. 4173-4179.
47. de Groot, F.M.F., et al., *2p x-ray absorption of 3d transition-metal compounds: An atomic multiplet description including the crystal field.* Physical Review B, 1990. **42**(9): p. 5459-5468.
48. Winterlik, J., et al., *Electronic, magnetic, and structural properties of the ferrimagnet Mn2CoSn.* Physical Review B, 2011. **83**(17).
49. Jang, Y., et al., *Magnetic field sensing scheme using CoFeB ∕ MgO ∕ CoFeB tunneling junction with superparamagnetic CoFeB layer.* Applied physics letters, 2006. **89**(16): p. 163119.
50. Magana, D., et al., *Switching-on superparamagnetism in Mn/CdSe quantum dots.* Journal of the American Chemical Society, 2006. **128**(9): p. 2931-2939.
51. Petracic, O., *Superparamagnetic nanoparticle ensembles.* Superlattices and Microstructures, 2010. **47**(5): p. 569-578.
52. Abbas, M., et al., *Size controlled sonochemical synthesis of highly crystalline superparamagnetic Mn–Zn ferrite nanoparticles in aqueous medium.* Journal of Alloys and Compounds, 2015. **644**: p. 774-782.
53. Köseoğlu, Y., et al., *Low temperature hydrothermal synthesis and characterization of Mn doped cobalt ferrite nanoparticles.* Ceramics International, 2012. **38**(5): p. 3625-3634.
54. Saranu, S., et al., *Effect of large mechanical stress on the magnetic properties of embedded Fe nanoparticles.* Beilstein Journal of Nanotechnology, 2011. **2**(1): p. 268-275.
55. Srivastava, M., M.B. Sahariah, and A. Srinivasan, *Size-dependent properties of single domain Fe 2CoGa nanoparticles prepared by a facile template-less chemical route.* Journal of Materials Chemistry C, 2022. **10**(33): p. 11946-11958.
56. Venugopalan, K., et al., *Magnetic properties of nano–sized Co2FeSi.* International journal of nanotechnology, 2011. **8**(10-12): p. 877-884.
57. Wang, C., et al., *Structural and magnetic properties of Fe2CoGa Heusler nanoparticles.* Journal of Physics D: Applied Physics, 2012. **45**(29): p. 295001.